\documentclass[%
 reprint,
 amsmath,amssymb,
 aps,
]{revtex4-2}
\usepackage[bbgreekl]{mathbbol}
\DeclareSymbolFontAlphabet{\mathbbl}{bbold}
\usepackage{amsfonts,amssymb,amsthm,mathtools}    
\DeclareSymbolFontAlphabet{\mathbbm}{bbold}
\DeclareSymbolFontAlphabet{\mathbb}{AMSb}%
\usepackage{graphicx}
\usepackage{dcolumn}
\usepackage{bm}
\usepackage{mathrsfs}
\usepackage[usenames]{color}\usepackage[dvipsnames]{xcolor}    
\usepackage{accents} 
\usepackage{tensor}
\usepackage{tikz}
\usepackage[breaklinks=true,backref=none]{hyperref} 

\newcommand{\corurl}{BrickRed}  \newcommand{\corcite}{red}
\newcommand{\corlink}{blue}    \newcommand{\corfile}{black}

\hypersetup{
	linktocpage,
	colorlinks,
	urlcolor=\corurl,	
	citecolor=\corcite,
	linkcolor=\corlink,
	filecolor=\corfile,
	pdfnewwindow,
	pdftitle={Equivalence symplectic forms},
	pdfauthor={JMB ESV},
	pdfsubject={CPS}}
\usetikzlibrary{matrix}

\def\ledgee{{\setbox0\hbox{\ensuremath{\mathrel{\cdot}}}\rlap{\hbox to \wd0{\hss\ensuremath\wedge\hss}}\box0}}

\usepackage{graphicx}
\makeatletter
\newcommand\rrule[3][0pt]{%
	\ifdim#2>#3\math@hrule[#1]{#2}{#3}\else\math@vrule[#1]{#2}{#3}\fi}
\newcommand\math@hrule[3][0pt]{%
	\gdef\mystery@factor{0.07}%
	\@tempdima=#3%
	\rule[#1]{0pt}{#3}
	\raisebox{.5\@tempdima+#1}{%
		\makebox[#2][l]{\kern-.5\@tempdima\@@mathrule{#2}{#3}}}%
}
\newcommand\math@vrule[3][0pt]{%
	\gdef\mystery@factor{0.0}%
	\@tempdima=#2%
	\rule[#1]{0pt}{#3}
	\raisebox{-.0\@tempdima+#1}{%
		\kern0.5\@tempdima%
		\rotatebox{90}{\kern-0.5\@tempdima\makebox[#3][l]{\@@mathrule{#3}{#2}}}%
		\kern0.5\@tempdima}%
}
\def\@@mathrule#1#2{%
	\@tempdimb=#2%
	\@tempdima=\dimexpr#1-\mystery@factor\@tempdimb
	\pdfliteral{%
		q []0 d %
		1 J 
		\strip@pt\@tempdimb\space w \strip@pt\@tempdimb\space 0 m %
		\strip@pt\@tempdima\space 0 l S Q }}
\makeatother

\graphicspath{{./font/}{./}}

\def\wwedgee{{\setbox0\hbox{\ensuremath{\mathrel{\wedge}}}\rlap{\hbox to \wd0{\hss\,\ensuremath\wedge\hss}}\box0}}

\newcommand{\wwedge}{\mathrel{\!\wwedgee\!}}
\def\ledgee{{\setbox0\hbox{\ensuremath{\mathrel{\cdot}}}\rlap{\hbox to \wd0{\hss\ensuremath\wedge\hss}}\box0}}

\newcommand{\dd}{\mathchoice
	{\mathbbm{d}\rrule{.087ex}{1.605ex}\hspace*{0.15ex}} 
	{\mathbbm{d}\rrule{.087ex}{1.605ex}\hspace*{0.15ex}} 
	{\mathbbm{d}\rrule{.08ex}{1.125ex}\hspace*{0.15ex}}  
	{\mathbbm{d}\rrule{.06ex}{.8ex}\hspace*{0.15ex}}     
}

\newlength{\alturaL}
\newcommand{\LL}{\includegraphics[height=1.1\alturaL]{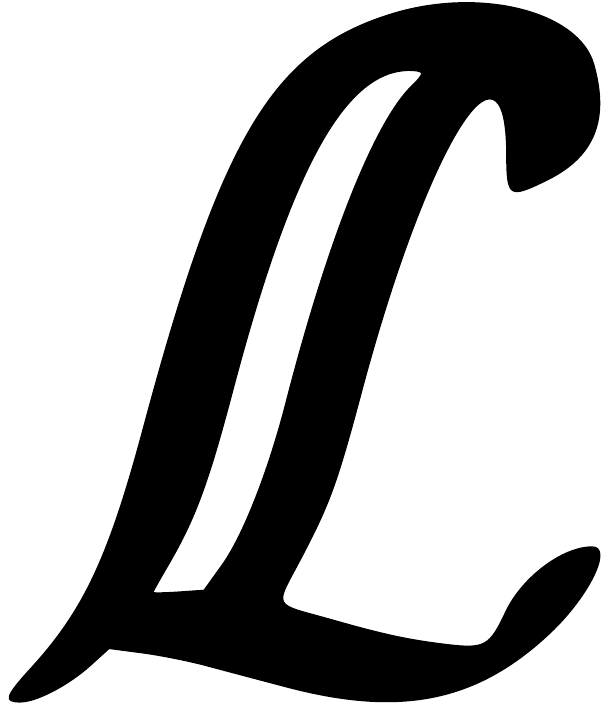}}

\newlength{\alturaO}
\newcommand{\OOmega}{\includegraphics[height=\alturaO]{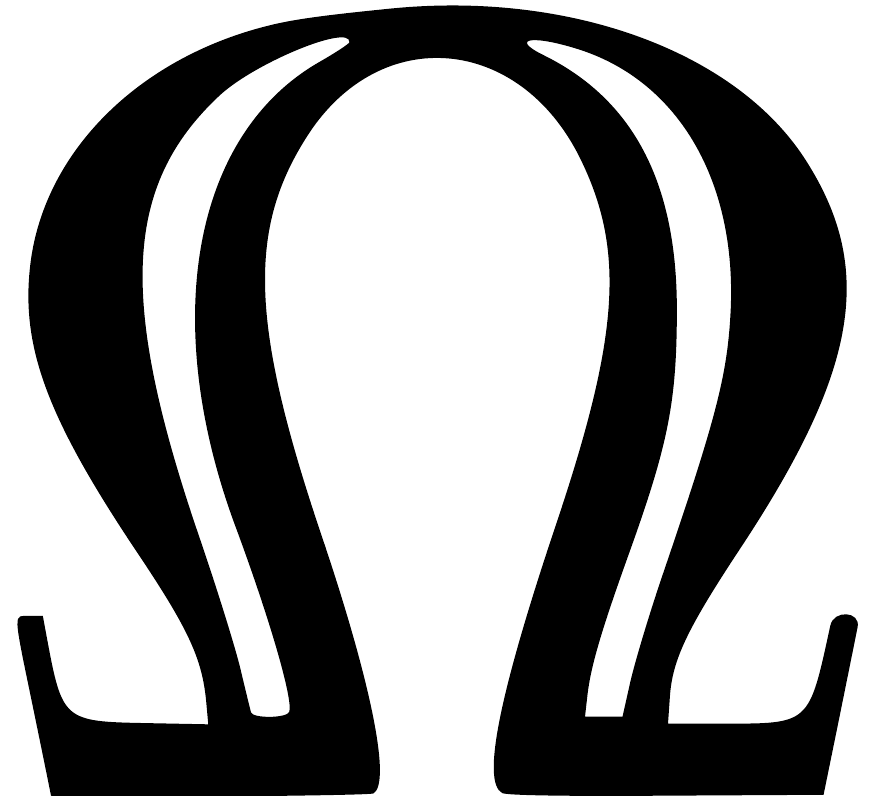}}

\newlength{\alturaI}
\newcommand{\ii}{\raisebox{-.04ex}{\includegraphics[height=1.1\alturaI]{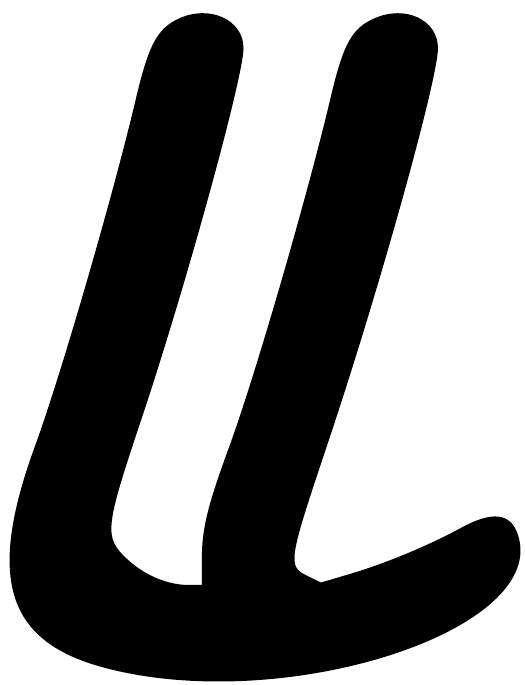}}}

\newcommand{\ele}{l}

\newlength{\alturaJ}
\newcommand{\jj}{\raisebox{-.37ex}{\includegraphics[height=.9\alturaJ]{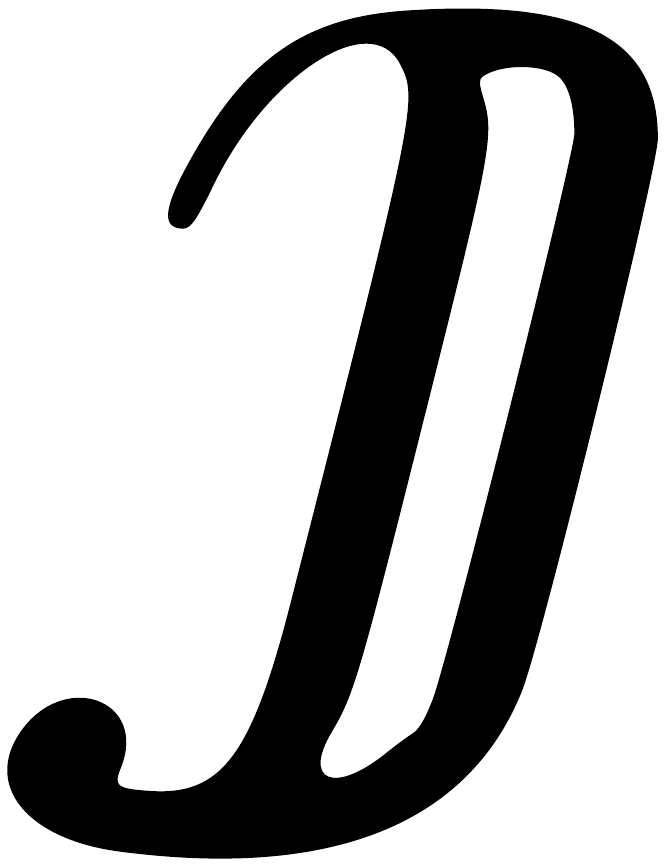}}}

\newlength{\alturaX}

\renewcommand{\L}{\mathcal{L}}

\newcommand{\VV}{\mathbb{V}}
\renewcommand{\SS}{\mathbb{S}}
\newcommand{\Sol}{\mathrm{Sol}}

\newcommand{\vol}{\mathrm{vol}}
\newcommand{\R}{\mathbb{R}}
\renewcommand{\d}{\mathrm{d}}

\newcommand{\F}{\mathcal{F}}
\newcommand{\Q}{\mathcal{Q}}

\usepackage{titlesec}
\usepackage{slashed}

\titlespacing\section{0pt}{10pt plus 4pt minus 2pt}{1pt plus 2pt minus 2pt}
\titlespacing\subsection{0pt}{12pt plus 4pt minus 2pt}{1pt plus 2pt minus 2pt}
\titlespacing\subsubsection{0pt}{12pt plus 4pt minus 2pt}{0pt plus 2pt minus 2pt}

\begin{document}

\preprint{APS/123-QED}

\title{Proof of the equivalence of the symplectic forms derived from the canonical and the covariant phase space formalisms}

\author{Juan Margalef-Bentabol$^{1,3}$}
\author{Eduardo J.S.~Villaseñor$^{2,3}$}
\affiliation{\rule{0ex}{3ex}${}^1$Department of Mathematics and Statistics, Memorial University, St. John's, Newfoundland and Labrador A1C 5S7, Canada.\\${}^2$Departamento de Matemáticas, Universidad Carlos III de Madrid. Avda. de la Universidad 30, 28911 Leganés, Spain.\\${}^3$Grupo de Teorías de Campos y Física Estadística. Instituto Gregorio Millán (UC3M). Unidad Asociada al Instituto de Estructura de la Materia, CSIC, Madrid, Spain.}

\begin{abstract}\noindent
 We prove that, for any theory defined over a space-time with boundary, the symplectic form derived in the covariant phase space is equivalent to the one derived from the canonical formalism.
\end{abstract}

\maketitle


\section{INTRODUCTION}\label{sec:INTRODUCTION}
In physics, we can identify, roughly speaking, two somewhat disjoint frameworks to deal with a field theory: the canonical and the covariant. The former breaks all the objects of the theory into spatial and normal components. The main advantages are that we gain a dynamical perspective and some universal structures like the symplectic form of a cotangent bundle. This provides a way to approach the problem numerically \cite{gourgoulhon20123+} (essential in the study of gravitational waves and LIGO observations) and a starting point for the Hamiltonian quantization \cite{woodhouse1997geometric,ashtekar2004background,margalef2018,juarez2015quantization}. A price to pay is the apparent loss of some symmetries.\vspace*{1.7ex}

The covariant approach, on the other hand, considers fields over the whole space-time \cite{gotay1998momentum}. Some noteworthy advantages are: symmetries are explicit, the study of null infinity is easier, there are methods to compute conserved quantities, and higher-derivative theories are treated on equal footing as 1st-order ones \cite{Stelle:1976gc,Stelle:1977ry,Lu:2015cqa}. All this has important consequences in effective and perturbation theory, both of which are relevant to the of study string theories, edge modes, corner and BMS algebras, or the analysis of consistent deformations \cite{Barnich:1993vg,Barnich:1994db,Barnich:1994mt,Odak:2021axr,Freidel:2021cjp,Freidel:2020xyx,Wieland:2021vef,Wieland:2020gno,concise,diaz2021hamiltonian}. However, one drawback is that there are no known canonical structures on the spaces involved. In particular, to have a symplectic structure, one has to rely on the covariant phase space (CPS) formalism and fix a local action.\vspace*{1.7ex}

The question that arises naturally is if both approaches are equivalent. It is possible to prove that in many aspects they are. However, the equivalence of the symplectic forms was unknown in general. The problem has been studied in concrete relevant examples, like GR without boundaries \cite{FrauendienerSparling1992,AshtekarMagnon1982}. Also in \cite{barnich1991covariant}, an analytical prove of the symplectic equivalence is provided over the reduced phase space (using Poisson brackets) for 1st-order theories with no boundaries. It is important to mention that adding boundaries complicates the matter a great deal even in concrete examples. In that sense, our paper \cite{CPS} was a breakthrough because it provided a way to map any theory with boundary to another one without boundary in the CPS formalism. Although this approach would simplify the upcoming computations, we will not use it here as it requires the introduction of a lot of notation and definitions. However, in \cite{CPS} we also provided a geometric language that bridges between the mathematical formalism ($\infty$-jets framework) and the standard physics notation. We will see in this paper that this language turns out to be essential  to prove, in full generality, the equivalence of both symplectic structures. As a byproduct, the equivalence shows that our proposal  of a CPS symplectic structure in manifolds with boundaries (introduced also in \cite{CPS}) is the most natural one. Not only for its cohomological nature, as explained in that paper, but also because it is equivalent to the one coming from the canonical formalism.

\section{THE GEOMETRIC ARENA}
\subsection{The spacetime}
Consider a globally hyperbolic $n$-manifold $M$ (up to diffeomorphism, $M=\R\times\Sigma$) with boundary  $\partial M=\R\times\partial\Sigma$. We have the inclusions $\overline{\jmath}:\partial\Sigma\hookrightarrow \Sigma$, $\jmath:\partial M\hookrightarrow M$, and $\imath_t:\Sigma\hookrightarrow \{t\}\times\Sigma\subset M$. As usual, we have the exterior derivative $\d$, the wedge product $\wedge$, the interior derivative $\iota_{\vec{V}}$, and the Lie derivative $\L_{\vec{V}}$.\vspace*{1.7ex}

In this setting it is crucial to keep track of the orientations. We orient $\R$ with the standard volume form $\d t$ and $\Sigma$ with some volume form $\vol_\Sigma$ (which, up to easy to handle technicalities, can be equally understood as an $(n-1)$-form on $\Sigma$ or on $M$). We then orient $M$ with
\begin{equation}\label{eq: vol M}
    \vol_M:=\d t\wedge\vol_\Sigma
\end{equation}
Boundaries are oriented so that Stokes' theorem holds
\begin{equation}\label{Stokes' theorem}
    \int_M \d\alpha=\int_{\partial M}\jmath^*\alpha\qquad\qquad\int_\Sigma \d\beta=\int_{\partial \Sigma}\overline{\jmath}^*\beta
\end{equation}
For that, take any metric $\gamma_{ij}$ on $\Sigma$ and denote $\nu^i$ the unit vector field $\gamma$-normal to $\partial\Sigma$. Consider the adapted metric $g_{\alpha\beta}$ on $\R\times\Sigma$ such that $g^{\alpha\beta}=-\partial_t^\alpha\partial_t^\beta+(\imath_t)^\alpha_i(\imath_t)^\beta_j\gamma^{ij}$, then the unit vector field $g$-normal to $\partial M$ is (up to pushforward) $\nu^i$, i.e., $\mathcal{V}^\alpha:=(\imath_t)^\alpha_i\nu^i$. Finally, we define
\begin{equation}
    \vol_{\partial \Sigma}:=\iota_{\vec{\nu}}\vol_\Sigma\qquad\qquad\vol_{\partial M}:=\iota_{\vec{\mathcal{V}}}\vol_M
\end{equation}
It is important to notice that
\begin{equation}\label{eq: vol partial M}
    \vol_{\partial M}=\iota_{\vec{\mathcal{V}}}(\d t\wedge \vol_\Sigma)=-\d t\wedge \vol_{\partial\Sigma}
\end{equation}

\subsection{The space of fields}

Let $\F$ be a space of tensor fields (of any tensorial character) over $M$. This space is $\infty$-dimensional and non-linear in general. Although it can rigorously be described with the $\infty$-jets formalism, for our purposes it is enough to think of $\F$ as a standard smooth manifold with the usual operators such as the exterior derivative $\dd$, the wedge product $\wwedge$\,, the interior derivative $\ii_{\VV}$, or the Lie derivative $\LL_{\VV}$. Here, $\VV$ is a vector field of $\F$ (see  \cite{CPS} for a careful discussion). Of course, $\F$ may consist of different types of tensor fields, hence $\F=\F^1\times\cdots\times\F^N$ with the fields labeled as $(\phi^I)_{I=1\cdots N}\in\F$.\vspace*{1.7ex}

\section{THE COVARIANT PHASE SPACE FORMALISM IN A NUTSHELL}
Roughly speaking, the CPS method studies the space of solutions of a theory over the whole space-time. The equations of motion are not dynamic but, rather, give conditions for the fields to be solutions. We devote this section to summarizing how to define a presymplectic form canonically associated with a local action. For a detailed discussion and some applications see \cite{CPS,CPSGR,CPSPT,CPSHolst}.\vspace*{1.7ex}

Consider a local action $\SS:\F\to\R$ given by
\begin{equation}\label{eq: S CPS}
\SS=\int_M L-\int_{\partial M}\overline{\ell}
\end{equation}
 where $(L,\overline{\ell})\in\Omega^{(n,0)}(M\times\F)\times\Omega^{(n-1,0)}(\partial M\times\F)$, known as CPS bulk and boundary Lagrangians, are top-forms on $(M,\partial M)$ and $0$-forms on $\F$ (they are bigraded forms). We assume that they are locally constructed, i.e., when evaluating $(L(\phi),\overline{\ell}(\phi))\in\Omega^n(M)\times\Omega^{n-1}(\partial M)$ at $p\in M$, they only depend on $p$, $\phi(p)$, and finitely many of its derivatives at $p$.\vspace*{1.7ex}

 It is a standard result \cite{CPS,anderson1989variational} that the $(n,1)$-form $\dd L$ can be split as
 \begin{equation}\label{eq: dd L}
 \dd L=E_I\wwedge\dd\phi^I+\d\Theta
 \end{equation}
 for some $(n,0)$-forms $E_I$ (Euler-Lagrange forms) and some $(n-1,1)$-form $\Theta$ (bulk symplectic potential, uniquely defined up to an exact form). In practice, this is achieved by using Leibniz's rule to remove all the derivatives from $\dd\phi^I$. Taking the $\dd$-exterior derivative of $\SS$, using \eqref{eq: dd L}, and Stokes' theorem, we have
\begin{equation}\label{eq: dd SS}
    \dd\SS=\int_M E^I\wwedge\dd\phi^I-\int_{\partial M}(\dd \overline{\ell}-\jmath^*\Theta)
\end{equation}
 We can split $\dd\overline{\ell}$ as in \eqref{eq: dd L}, but we have the additional term $\jmath^*\Theta$ so we need it to be ``splittable'' as well. This condition has to be imposed, in which case we say that $\SS$  defines a good variational principle. This leads to
  \begin{equation}\label{eq: dd ell}
 \dd \overline{\ell}-\jmath^*\Theta=\overline{b}_I\wwedge\dd\phi^I-\d\overline{\theta}
 \end{equation}
 for some $(n-1,0)$-forms $\overline{b}_I$ (boundary Euler-Lagrange forms) and some $(n-2,1)$-form $\overline{\theta}$ (boundary symplectic potential, uniquely defined up to an exact form). With these ingredients, we define the space of solutions
 \[\Sol(\SS)=\{\phi\in\F\ /\ (E^I(\phi),\overline{b}^I(\phi))=(0,0)\}\overset{\jj_\SS}{\hookrightarrow}\F\]
 and the symplectic structure associated with an embedding $\imath:\Sigma\hookrightarrow M$
 \begin{equation}\label{eq: OOmega SS}
     \OOmega_\SS^\imath=\int_\Sigma\dd\imath^*\Theta-\int_{\partial\Sigma}\dd\imath^*\overline{\theta}
 \end{equation}
$\OOmega_\SS^\imath$ is independent of the choice of Lagrangians (as long as they define the same $\SS$) and of the symplectic potentials chosen in \eqref{eq: dd L} and \eqref{eq: dd ell}. Moreover, if we denote $\OOmega_\SS:=\jj_\SS^*\Omega_\SS^\imath$ the pullback of the symplectic form to $\Sol(\SS)$, then it can be proved that $\OOmega_\SS$ does not depend on the embedding either. Thus, we have constructed a presymplectic form on $\Sol(\SS)$ canonically associated with $\SS$.

 \section{THE CANONICAL FORMALISM IN A NUTSHELL}
The canonical (CAN) formalism  deals, roughly speaking, with ``instant fields'' that evolve according to some dynamical equations. By evolving specific initial data, we obtain a curve in the space of ``instantaneous fields'' which corresponds (up to integrability issues) to a solution over the whole space-time. We devote this section to summarize the results necessary for the present work. For now, we will focus on 1st-order theories and delay the generalization to higher order ones to section \ref{section: higher order}.\vspace*{1.7ex}

Consider the space $\Q=\Q^1\times\cdots\times\Q^M$ of fields over $\Sigma$ and its tangent bundle $T\Q$ (its standard geometric operators like $\dd$ or $\wwedge$ are denoted as the ones of $\F$). It consists of elements of the form $(q^1,\ldots,q^M;v^1,\ldots,v^M)$. Consider a local CAN Lagrangian-action $\mathcal{L}:T\Q\to\R$ given by
\begin{equation}\label{eq: canonical lagrangian}
    \mathcal{L}=\int_\Sigma \mathrm{L}-\int_{\partial\Sigma}\ele
\end{equation}
where $(\mathrm{L},\ele)\in\Omega^{(n-1,0)}(\Sigma\times T\Q)\times\Omega^{(n-2,0)}(\partial \Sigma\times T\Q)$ are some (locally constructed and possibly time-dependent) CAN Lagrangians. $\mathcal{L}$ is historically called Lagrangian but it plays a role similar to the action \eqref{eq: S CPS} although not entirely equal since the time integration is missing. So we define the CAN action $\mathcal{S}:\mathcal{C}^\infty(\R,\Q)\to\R$
\begin{equation}\label{eq: S(q)}
\mathcal{S}(q)=\int_\R\mathcal{L}\big(q(t),\dot{q}(t)\big)\d t
\end{equation}

The analogues to equations \eqref{eq: dd L} and \eqref{eq: dd ell} are
\begin{align}\begin{split}\label{eq: dd L canonic}
    &\dd \mathrm{L}=A_I^{(0)}\wwedge\dd q^I+A_I^{(1)}\wwedge\dd v^I+\d\widetilde{\Theta}\\
    &\dd \ele-\overline{\jmath}^*\widetilde{\Theta}=B_I^{(0)}\wwedge\dd q^I+B_I^{(1)}\wwedge\dd v^I-\d\widetilde{\theta}
\end{split}\end{align}
Taking the $\dd$-exterior derivative of \eqref{eq: S(q)}, using Stokes' theorem, and integrating by parts with respect to time (notice that in $\mathcal{S}$ every $v^I$ is replaced by $\dot{q}^I$), we obtain
\begin{align*}
    \dd\mathcal{S}\!=\!\!\int_\R\!\d t\Big(\!\int_\Sigma(A_I^{(0)}-\dot{A}_I^{(1)})\!\wwedge\dd q^I\!\!-\!\!\int_{\partial\Sigma}(B_I^{(0)}-\dot{B}_I^{(1)})\!\wwedge\dd q^I\!\Big)
\end{align*}
$(A_I^{(0)}-\dot{A}_I^{(1)},B_I^{(0)}-\dot{B}_I^{(1)})$ are the bulk and boundary dynamical equations (if some of them do not involve time derivatives, we obtain bulk or boundary constraints).\vspace*{1.7ex}

Once we have the Lagrangian formalism, we proceed to introduce the Hamiltonian formalism which, instead of living in the tangent bundle $T\Q$, lives on the cotangent bundle $T^*\!\Q$. The advantage of the latter is its canonical symplectic form, which plays an essential role on the Hamiltonian formulation. Indeed, denoting the elements of $T^*\!\Q$ as $(q;p)=(q^1,\ldots,q^M;p_1,\ldots,p_M)$, the canonical symplectic structure is given by
\begin{equation}\label{eq: canonical symplectic}
    \OOmega_{T^*\!\Q}:=\dd q^I\wwedge \dd p_I
\end{equation}

The usual pairing between position and momenta applies when evaluated over fields. In order to go from the Lagrangian (tangent bundle) to the Hamiltonian (cotangent bundle), we use the fiber derivative $F\mathcal{L}:T\Q\to T^*\!\Q$. For each $(q;v)\in T_q\Q$ we define     $F\mathcal{L}_{(q;v)}\in T_q^*\Q$ as
\begin{equation}
    F\mathcal{L}_{(q;v)}\big(q;w\big)=\left.\frac{\d}{\d\tau}\right|_{\tau=0}\mathcal{L}\big(q;v_\tau\big)
\end{equation}
where $v_\tau$ is a curve in $T_q\Q$ with $v_0=v$ and $\left.\frac{\d}{\d \tau}\right|_0v_\tau=w$. In general, this map is not surjective and the relevant symplectic structure is actually the one induced on its image (the primary constraint submanifold), i.e., induced by the inclusion $\jj_{\mathcal{L}}:F\mathcal{L}(T\Q)\hookrightarrow T^*\!\Q$. Since over $F\mathcal{L}(T\Q)$ we have $p=F\mathcal{L}_{(q;v)}$, we schematically obtain
\begin{equation}\label{eq: OOmega second try}
    \OOmega_{\mathcal{L}}:=\jj_{\mathcal{L}}^* \OOmega_{T^*\!\Q}=\dd q^I\wwedge \dd \jj_{\mathcal{L}}^*p_I=\dd q^I\wwedge \dd  (F\mathcal{L}_{(q;v^I)})
\end{equation}
where by $v^I$ we mean $(0,\ldots,0,v^I,0,\ldots,0)$. In order to give a concrete sense to \eqref{eq: OOmega second try}, we rely on the geometric language introduced before. The key observation is that the momenta can be rewritten as
\begin{align}\begin{split}\label{def p}
    &p_I(q;w^I):=\LL_{(0,w^I)}\mathcal{L}=\ii_{(0,w^I)}\dd\mathcal{L}=\\
    &=\int_\Sigma \ii_{(0,w^I)}\dd\mathrm{L}-\int_{\partial\Sigma}\ii_{(0,w^I)}\dd\ele
\end{split}
\end{align}
The Lie derivative acts following Cartan's rule while the interior derivative acts as follows: $\ii_{(\alpha^I,\beta^J)}\dd q^K=\delta^K_I \alpha^I$ and $\ii_{(\alpha^I,\beta^J)}\dd v^K=\delta^K_J \beta^J$. Using \eqref{def p}, equation \eqref{eq: dd L canonic}, and Stokes' theorem leads to (removing the argument)
\begin{align*}
    p_I(\cdot{})=\int_\Sigma A_I^{(1)}\wwedge \cdot{}-\int_{\partial\Sigma}B_I^{(1)}\wwedge \cdot{}
\end{align*}
Taking its $\dd$-exterior derivative and using the $p-q$ pairing finally allows us to rewrite \eqref{eq: OOmega second try} as
\begin{equation}\label{eq: OOmega third try}
     \OOmega_{\mathcal{L}}=\int_\Sigma \dd A_I^{(1)}\wwedge \dd q^I-\int_{\partial\Sigma}\dd B_I^{(1)}\wwedge \dd q^I
\end{equation}

\section{PROVING THE EQUIVALENCE FOR FIRST ORDER THEORIES}\label{section: equivalence 1st}
In the previous sections we have seen that, on the one hand, we can derive a symplectic structure $\OOmega_\SS$ canonically associated with a CPS action $\SS$ (defined over the space of fields of $M$). On the other hand, from a CAN Lagrangian-action $\mathcal{L}$ defined over some $T\mathcal{Q}$, we get the symplectic structure $\OOmega_{\mathcal{L}}$ canonically associated with $\mathcal{L}$. It is well known that we can go from the CPS formalism to the CAN one by performing a $(1,n-1)$-decomposition (there are many equivalent ways of doing it). Therefore, for a given theory we  end up with two symplectic structures canonically associated with it. The issue of understanding the relation between them has been a long-lasting open question that we answer now.

\subsection{CANonicalizing the Lagrangian}\label{subsection: Canonicalizing the Lagrangian}
In order to go from the CPS Lagrangians $(L,\overline{\ell})$ to the CAN Lagrangians $(\mathrm{L},\ele)$, we need Fubini's theorem.

\begin{center}
\begin{minipage}{.455\textwidth}
\textbf{Theorem}\mbox{}\\
Let $A$ and $B$ be manifolds oriented with $\vol_A$ and $\vol_B$ respectively. We orient $A\times B$ with $\vol_A\wedge\vol_B$ and consider an integrable function $f:A\times B\to\R$, then
\[\int_{A\times B}f\vol_A\wedge\vol_B=\int_A\left(\int_B f\vol_B\right)\vol_A \]
\end{minipage}%
\end{center}%
Let us now use the $(1,n-1)$-decomposition induced from $M=\R\times\Sigma$ to decompose the CPS Lagrangians. In general, given $\alpha\in\Omega^k(M)$, we have $\alpha=\d t\wedge\alpha_\perp+\alpha^\top$ with $\alpha_\perp:=\iota_{\partial_t}\alpha$ and $\alpha^\top:=\iota_{\partial_t}(\d t\wedge\alpha)$. Since the CPS Lagrangians are top-forms, we simply have
    $L=\d t\wedge L_\perp$ and 
    $\overline{\ell}=\d t\wedge \overline{\ell}_\perp$.
Besides, there exist $F,f$ such that 
\begin{align*}
    &\d t\wedge L_\perp=L=F\vol_M\!\overset{\eqref{eq: vol M}}{=}\!\d t\wedge(F\vol_\Sigma)&&\!\!\!\!\!\!\to\ F=\frac{L_\perp}{\vol_\Sigma}\\
    &\d t\wedge \overline{\ell}_\perp=\overline{\ell}=f\vol_{\partial M}\!\overset{\eqref{eq: vol partial M}}{=}\!(-\d t)\wedge(f \vol_{\partial\Sigma})&&\!\!\!\!\!\!\to\ f=\frac{-\overline{\ell}_\perp}{\vol_{\partial\Sigma}}
\end{align*}
where $\frac{\omega}{\vol}$ denotes the function that relates the top-form $\omega$ with the volume form vol. Plugging these expressions of $(L,\overline{\ell})$ into \eqref{eq: S CPS} and using Fubini's theorem we get
\begin{align*}
\SS&=\int_\R\left(\int_{\Sigma} L_\perp-\int_{\partial\Sigma}(-\overline{\ell}_\perp)\right)\d t
\end{align*}
A minus sign appears in the boundary because, from \eqref{eq: vol partial M}, we have $(A,\vol_A)=(\R,-\d t)$ in Fubini's theorem.\vspace*{1.7ex}

All this allows us to identify $\mathrm{L}$ with $\iota_{\partial_t}L$ and $\ele$ with $-\iota_{\partial_t}\overline{\ell}$. To formalize this identification, we perform the $(1,n-1)$-decomposition on the fields $(\phi^I)_I$ and their derivatives. Since the theory is of 1st order, we end up with some tangential fields $(\hat{q}^J)_J$ and their velocities $\hat{v}^J:=\L_{\partial_t}\hat{q}^J$. We now express $(\iota_{\partial_t}L,-\iota_{\partial_t}\overline{\ell})$ in terms of $(\hat{q}^J;\hat{v}^J)_J$ and pull everything back to $\Sigma$ (where the positions and velocities are denoted as $(q^J;v^J)_J$) to obtain $(\mathrm{L},\ele)$. Plugging them into equation \eqref{eq: canonical lagrangian} leads to a CAN Lagrangian-action associated with $\SS$ that we denote $\mathcal{L}_\SS$.

\subsection{CANonicalizing the symplectic potentials}\label{subsection: Canonicalizing the Symplectic Potentials}
Now we want to rewrite the symplectic potentials derived from the CPS formalism in a way suitable for the comparison with the CAN formalism. For that, we compute
\begin{align*}
    &E_I\wwedge\dd\phi^I+\d\Theta=\dd L=\d t\wedge\dd \iota_{\partial_t} L\overset{\dagger}{=}\\
    &=\d t\wedge(\hat{A}_I^{(0)}\wwedge\dd \hat{q}^I+\hat{A}_I^{(1)}\wwedge\dd \L_{\partial_t}\hat{q}^I+\d\hat{\Theta})=\\
    &=\d t\wedge(\hat{A}_I^{(0)}-\L_{\partial_t}\hat{A}_I^{(1)})\wwedge\dd \hat{q}^I+\\
    &\qquad+\d\Big(\iota_{\partial_t}(\d t\wedge \hat{A}_I^{(1)}\wwedge\dd \hat{q}^I)-\d t\wedge\hat{\Theta}\Big)
\end{align*}
In the $\dagger$ equality we have ``lifted'' \eqref{eq: dd L canonic} from $\Sigma$ to $M$ (the corresponding objects are denoted with a hat) taking into account that, although additional terms might seem to appear, they have a $\d t$ in them so they vanish in our previous computation. We know from \cite{wald1990identically,CPS,takens1979global}, that if $r<n$ and $s>0$, then any $\d$-closed $(r,s)$-form is also $\d$-exact. So there exists some $Z\in\Omega^{(n-2,1)}(M)$ such that
\begin{equation}\label{eq: Theta CPC to canonical}
    \Theta=\iota_{\partial_t}(\d t\wedge \hat{A}_I^{(1)}\wwedge\dd \hat{q}^I)-\d t\wedge\hat{\Theta}+\d Z
\end{equation}
Since $\jmath^*$ and $\iota_{\partial_t}$ commute ($\partial_t$ is tangent to the boundary), if we compute $\jmath^*\Theta$, the first term vanishes. Hence
\begin{align*}
    &b_I\wwedge\dd\phi^I-\d\overline{\theta}=\dd\overline{\ell}-\jmath^*\Theta\overset{\eqref{eq: Theta CPC to canonical}}{=}\\
    &=-\d t\wedge(\dd(-\iota_{\partial_t}\overline{\ell})-\jmath^*\hat{\Theta})-\d\jmath^*Z\overset{\eqref{eq: dd L canonic}}{=}\\
    &=-\d t\wedge(\hat{B}_I^{(0)}-\L_{\partial_t}\hat{B}_I^{(1)})\wwedge\dd \hat{q}^I-\\
    &\qquad-\d\Big(\iota_{\partial_t}(\d t\wedge \hat{B}_I^{(1)}\wwedge\dd \hat{q}^I)-\d t\wedge\hat{\theta}+\jmath^*Z\Big)
\end{align*}
Thus, there exists some $\overline{z}\in\Omega^{(n-3,1)}(\partial M)$ such that
\begin{equation}
\overline{\theta}=\iota_{\partial_t}(\d t\wedge \hat{B}_I^{(1)}\wwedge\dd \hat{q}^I)-\d t\wedge\hat{\theta}+\jmath^*Z-\d\overline{z}
\end{equation}
Finally, we rewrite  the CPS symplectic potentials in a more suitable way:
\begin{align}\label{eq: Theta CPS=canonical}
    \begin{split}
        &\Theta=\hat{A}_I^{(1)}\wwedge\dd \hat{q}^I+\d t\wedge(\cdots)+\d Z\\
        &\overline{\theta}=\hat{B}_I^{(1)}\wwedge\dd \hat{q}^I+\d t\wedge (\cdots)-\d t\wedge\hat{\theta}+\jmath^*Z-\d\overline{z}
    \end{split}
\end{align}

\subsection{CANonicalizing the symplectic form} 
Since the symplectic structure over $\Sol(\SS)$ does not depend on the embedding, we consider $\imath_{t_0}$ (where $\d t=0$, so in particular $\imath_{t_0}^*\hat{A}_I=A_I$ and $\imath_{t_0}^*\hat{B}_I=B_I$) and we finally obtain the desired equivalence
\begin{align}\label{eq: equivalence}
    \begin{split}
        \OOmega_\SS\overset{\eqref{eq: OOmega SS}}{\underset{\eqref{eq: Theta CPS=canonical}}{=}}\int_\Sigma\dd A_I^{(1)}\wwedge\dd q^I-\int_{\partial\Sigma}\dd B_I^{(1)}\wwedge\dd q^I\overset{\eqref{eq: OOmega third try}}{=}  \OOmega_{\mathcal{L}_\SS}  
    \end{split}
\end{align}

\section{PROVING THE EQUIVALENCE FOR HIGHER ORDER THEORIES}\label{section: higher order}
In section \ref{subsection: Canonicalizing the Lagrangian} we obtained the 1st-order CAN Lagrangians from the CPS ones by breaking $(\phi^I)_I$ into $(\hat{q}^J,\L_{\partial_t}\hat{q}^J)_J$. However, for general theories higher order ``velocities'' $\{(\L_{\partial_t})^\mu\hat{q}^J\}_{\mu=1\cdots K}$ will appear. While the CPS formalism does not change, the CAN one changes drastically. We devote this section to briefly summarizing how to prove the equivalence for higher order theories.

 \subsection{The higher order canonical formalism in a nutshell}
For a detailed description, see \cite{de2011generalized}. For our purposes, we only need to generalize some of the equations of section \ref{subsection: Canonicalizing the Lagrangian}. First, equation \eqref{eq: dd L canonic} has to be changed to
\begin{align}\begin{split}\label{eq: dd L canonic higher}
    &\dd \mathrm{L}=\sum_{\mu}A_I^{(\mu)}\wwedge\dd q_{(\mu)}^I+\d\widetilde{\Theta}\\
    &\dd \ele-\overline{\jmath}^*\widetilde{\Theta}=\sum_{\mu}B_I^{(\mu)}\wwedge\dd q_{(\mu)}^I-\d\widetilde{\theta}
\end{split}\end{align}
where $q^I_{(0)}$ are the positions, $q^I_{(1)}$ their velocities, $q^I_{(2)}$ their accelerations, and so on up to $\mu=K$. Thus, we are working on the $K$-th tangent bundle $T^K\!\Q$. The Hamiltonian formalism takes place in $T^*\!(T^{K-1}\!\Q)$ where we have $\{q^I_{(\mu)}\}_{\mu=0\cdots K-1}$ and their momenta $\{p_I^{(\mu)}\}_{\mu=0\cdots K-1}$. The canonical symplectic structure generalizing \eqref{eq: canonical symplectic} is 
\begin{equation}\label{eq: canonical symplectic higher}
    \OOmega_{T^*\!(T^{K-1}\Q)}:=\sum_{\mu=0}^{K-1}\dd q^I_{(\mu)}\wwedge \dd p_I^{(\mu)}
\end{equation}

In order to go from the Lagrangian to the Hamiltonian formulation, we have to generalize the fiber derivative. For our purposes it is enough to generalize \eqref{def p}. The standard way is
\begin{align*}
    &p^{(K)}_I(q;w^I):=\LL_{(0,w^I)}\mathcal{L}\\
    &p^{(\mu)}_I(q;w^I):=\LL_{(0,w^I)}\mathcal{L}-\frac{\d}{\d t}p^{(\mu+1)}_I(q;w^I)
\end{align*}
for $\mu=1,\ldots,K-1$. It is not hard to prove by (backwards) induction that
\begin{align*}
    p_I^{(\mu)}(\cdot{})=\sum_{k=\mu}^K\left(\!-\frac{\d}{\d t}\right)^{\!k-\mu}\left\{\int_\Sigma A_I^{(k)}\wwedge \cdot{}-\int_{\partial\Sigma}B_I^{(k)}\wwedge \cdot{}\right\}
\end{align*}
Taking the $\dd$-exterior derivative, plugging the result into \eqref{eq: canonical symplectic higher}, and performing some standard manipulations with double finite sums, we obtain
\begin{align}\begin{split}\label{eq: OOmega canonical higher}
     &\OOmega_{\mathcal{L}}=\sum_{k=1}^K\sum_{\mu=0}^{k-1}\left\{\int_\Sigma \dd\!\left(\!-\frac{\d}{\d t}\right)^{\!k-1-\mu}\!A_I^{(k)}\wwedge \dd q^I_{(\mu)}-\right.\\
     &\qquad\qquad\left.-\int_{\partial\Sigma}\dd\!\left(\!-\frac{\d}{\d t}\right)^{\!k-1-\mu}\!B_I^{(k)}\wwedge \dd q^I_{(\mu)}\right\}
    \end{split}
\end{align}
Taking $K=1$ leads, as expected, to \eqref{eq: OOmega third try}.

\subsection{Equivalence for higher order theories}

Reasoning by induction, using computations similar to those of section \ref{subsection: Canonicalizing the Symplectic Potentials}, and equation \eqref{eq: dd L canonic higher} (where $q^I_{(\mu)}$ has to be replaced by $(\L_{\partial_t})^\mu \hat{q}^I$ since we are in the CPS formalism), we obtain
\begin{align}
    \begin{split}
        &\dd L=\d t \wedge\sum_\mu(-\L_{\partial_t})^\mu \hat{A}_I^{(\mu)}\wedge\dd \hat{q}^I+\d\Theta\\
        &\dd\overline{\ell}-\jmath^*\Theta=-\d t\wedge\sum_\mu(-\L_{\partial_t})^\mu\hat{B}_I^{(\mu)}\wedge\dd \hat{q}^I-\d\overline{\theta}
    \end{split}
\end{align}
where (up to exact forms)
\begin{align*}
       &\Theta=\sum_{k=1}^K\sum_{\mu=0}^{k-1}(-\L_{\partial_t})^{k-1-\mu}\hat{A}_I^{(k)}\wedge\dd (\L_{\partial_t})^\mu \hat{q}^I-\d t\wedge (\cdots)\\
       &\overline{\theta}=\sum_{k=1}^K\sum_{\mu=0}^{k-1}(-\L_{\partial_t})^{k-1-\mu}\hat{B}_I^{(k)}\wedge\dd (\L_{\partial_t})^\mu \hat{q}^I-\d t\wedge (\cdots)
\end{align*}
Plugging them into \eqref{eq: OOmega SS}, considering an embedding $\imath_{t_0}$, and taking into account that in CAN $(\L_{\partial_t})^\mu \hat{q}^I$ is identified with $q^I_{(\mu)}$, we obtain the general equivalence
\[\OOmega_\SS\overset{\eqref{eq: OOmega canonical higher}}{=}\OOmega_{\mathcal{L}_\SS}\]

\section{CONCLUSIONS AND COMMENTS}\label{sec:CONCLUSIONS}
We have proved the equivalence of the symplectic form induced by the Hamiltonian formalism and the one derived in the covariant phase space. The equivalence, which has been an open question for several decades, holds for theories of any order and even when boundaries are present. The proof relies strongly on the geometric formalism introduced in \cite{CPS}, where we also checked the equivalence in some concrete examples. Finally, this work also proves that the symplectic form introduced in \cite{CPS} for manifolds with boundaries is the natural one.

\begin{acknowledgments}
This work has been supported by the Spanish Ministerio de Ciencia e Innovaci\'on-AEI PID2020-116567GB-C22 grants. J.M.B. is supported by the AARMS postdoctoral fellowship and by the NSERC Grants Discovery-2018-04873 and RGPIN-2018-04887. E.J.S.V. is supported by the Madrid Government (Comunidad de Madrid-Spain) under the Multiannual Agreement with UC3M in the line of Excellence of University Professors (EPUC3M23), and in the context of the V PRICIT (Regional Programme of Research and Technological Innovation). We thank F. Barbero for his very useful comments and discussions.
\end{acknowledgments}
 
 \vspace*{-3ex}

 \setlength{\labelsep}{.25ex}%

	\bibliographystyle{plainnat}
\bibliography{bibliography}

\begin{thebibliography}{30}
\providecommand{\natexlab}[1]{#1}
\providecommand{\url}[1]{\texttt{#1}}
\expandafter\ifx\csname urlstyle\endcsname\relax
  \providecommand{\doi}[1]{doi: #1}\else
  \providecommand{\doi}{doi: \begingroup \urlstyle{rm}\Url}\fi

\bibitem[Gourgoulhon(2012)]{gourgoulhon20123+}
E.~Gourgoulhon.
\newblock \emph{3+1 formalism in general relativity: bases of numerical
  relativity}, volume 846.
\newblock
  \href{https://link.springer.com/book/10.1007/978-3-642-24525-1}{Springer
  Science \& Business Media}, 2012.

\bibitem[Woodhouse(1997)]{woodhouse1997geometric}
N.M.J. Woodhouse.
\newblock \emph{Geometric quantization}.
\newblock
  \href{https://global.oup.com/academic/product/geometric-quantization-9780198502708?cc=ca&lang=en&}{Oxford
  university press}, 1997.

\bibitem[Ashtekar and Lewandowski(2004)]{ashtekar2004background}
A.~Ashtekar and J.~Lewandowski.
\newblock \emph{Background independent quantum gravity: A status report}.
\newblock \href{http://dx.doi.org/10.1088/0264-9381/21/15/R01}{Class. Quant.
  Grav.}, 21\penalty0 (15):\penalty0 R53, 2004.
\newblock [arXiv:\href{http://arxiv.org/abs/gr-qc/0404018}{gr-qc/0404018}].

\bibitem[Margalef-Bentabol(2018)]{margalef2018}
J.~Margalef-Bentabol.
\newblock \emph{Towards general relativity through parametrized theories}.
\newblock PhD thesis, Universidad Carlos III de Madrid, 2018.
\newblock [arXiv:\href{http://arxiv.org/abs/1807.05534}{1807.05534}].

\bibitem[Ju{\'a}rez-Aubry et~al.(2015)Ju{\'a}rez-Aubry, Barbero~G.,
  Margalef-Bentabol, and Villase{\~n}or]{juarez2015quantization}
B.A. Ju{\'a}rez-Aubry, J.F. Barbero~G., J.~Margalef-Bentabol, and E.J.
  Villase{\~n}or.
\newblock \emph{Quantization of scalar fields coupled to point masses}.
\newblock Classical and Quantum Gravity, 32\penalty0 (24):\penalty0 245009,
  2015.

\bibitem[Gotay et~al.(1998)Gotay, Isenberg, Marsden, and
  Montgomery]{gotay1998momentum}
M.J. Gotay, J.~Isenberg, J.E. Marsden, and R.~Montgomery.
\newblock \emph{Momentum maps and classical relativistic fields. part {I}:
  Covariant field theory}.
\newblock Preprint, 1998.
\newblock [arXiv:\href{http://arxiv.org/abs/physics/9801019}{physics/9801019}].

\bibitem[Stelle(1977)]{Stelle:1976gc}
K.S. Stelle.
\newblock \emph{Renormalization of higher derivative quantum gravity}.
\newblock \href{http://dx.doi.org/10.1103/PhysRevD.16.953}{Phys. Rev. D},
  16:\penalty0 953, 1977.

\bibitem[Stelle(1978)]{Stelle:1977ry}
\rule[.6ex]{3em}{.05ex}.
\newblock \emph{Classical gravity with higher derivatives}.
\newblock \href{http://dx.doi.org/10.1007/BF00760427}{Gen. Rel. Grav.},
  9:\penalty0 353, 1978.

\bibitem[Lu et~al.(2015)Lu, Perkins, Pope, and Stelle]{Lu:2015cqa}
H.~Lu, A.~Perkins, C.N. Pope, and K.S. Stelle.
\newblock \emph{Black holes in higher-derivative gravity}.
\newblock \href{http://dx.doi.org/10.1103/PhysRevLett.114.171601}{Phys. Rev.
  Lett.}, 114\penalty0 (17):\penalty0 171601, 2015.
\newblock [arXiv:\href{http://arxiv.org/abs/1502.01028}{1502.01028}].

\bibitem[Barnich and Henneaux(1993)]{Barnich:1993vg}
G.~Barnich and M.~Henneaux.
\newblock \emph{Consistent couplings between fields with a gauge freedom and
  deformations of the master equation}.
\newblock \href{http://dx.doi.org/10.1016/0370-2693(93)90544-R}{Phys. Lett. B},
  311:\penalty0 123, 1993.
\newblock [arXiv:\href{http://arxiv.org/abs/hep-th/9304057}{hep-th/9304057}].

\bibitem[Barnich et~al.(1995{\natexlab{a}})Barnich, Brandt, and
  Henneaux]{Barnich:1994db}
G.~Barnich, F.~Brandt, and M.~Henneaux.
\newblock \emph{Local {BRST} cohomology in the antifield formalism: {I}.
  {G}eneral theorems}.
\newblock \href{http://dx.doi.org/10.1007/BF02099464}{Commun. Math. Phys.},
  174:\penalty0 57, 1995{\natexlab{a}}.
\newblock [arXiv:\href{http://arxiv.org/abs/hep-th/9405109}{hep-th/9405109}].

\bibitem[Barnich et~al.(1995{\natexlab{b}})Barnich, Brandt, and
  Henneaux]{Barnich:1994mt}
\rule[.6ex]{3em}{.05ex}.
\newblock \emph{Local {BRST} cohomology in the antifield formalism: {II}.
  {A}pplication to {Y}ang-{M}ills theory}.
\newblock \href{http://dx.doi.org/10.1007/BF02099465}{Commun. Math. Phys.},
  174:\penalty0 93, 1995{\natexlab{b}}.
\newblock [arXiv:\href{http://arxiv.org/abs/hep-th/9405109}{hep-th/9405109}].

\bibitem[Odak and Speziale(2021)]{Odak:2021axr}
G.~Odak and S.~Speziale.
\newblock \emph{Brown-{Y}ork charges with mixed boundary conditions}.
\newblock \href{http://dx.doi.org/10.1007/JHEP11(2021)224}{JHEP}, 11\penalty0
  (11):\penalty0 224, 2021.
\newblock [arXiv:\href{http://arxiv.org/abs/2109.02883}{2109.02883}].

\bibitem[Freidel et~al.(2021)Freidel, Oliveri, Pranzetti, and
  Speziale]{Freidel:2021cjp}
L.~Freidel, R.~Oliveri, D.~Pranzetti, and S.~Speziale.
\newblock \emph{Extended corner symmetry, charge bracket and
  {E}instein\textquoteright{}s equations}.
\newblock \href{http://dx.doi.org/10.1007/JHEP09(2021)083}{JHEP}, 09\penalty0
  (083):\penalty0 083, 2021.
\newblock [arXiv:\href{http://arxiv.org/abs/2104.12881}{2104.12881}].

\bibitem[Freidel et~al.(2020)Freidel, Geiller, and Pranzetti]{Freidel:2020xyx}
L.~Freidel, M.~Geiller, and D.~Pranzetti.
\newblock \emph{Edge modes of gravity. {P}art {I}. {C}orner potentials and
  charges}.
\newblock \href{http://dx.doi.org/10.1007/JHEP11(2020)026}{JHEP}, 11\penalty0
  (11):\penalty0 026, 2020.
\newblock [arXiv:\href{http://arxiv.org/abs/2006.12527}{2006.12527}].

\bibitem[Wieland(2021{\natexlab{a}})]{Wieland:2021vef}
W.~Wieland.
\newblock \emph{Gravitational {SL}(2,{R}) algebra on the light cone}.
\newblock \href{http://dx.doi.org/10.1007/JHEP07(2021)057}{JHEP}, 07\penalty0
  (057):\penalty0 057, 2021{\natexlab{a}}.
\newblock [arXiv:\href{http://arxiv.org/abs/2104.05803}{2104.05803}].

\bibitem[Wieland(2021{\natexlab{b}})]{Wieland:2020gno}
\rule[.6ex]{3em}{.05ex}.
\newblock \emph{Null infinity as an open {H}amiltonian system}.
\newblock \href{http://dx.doi.org/10.1007/JHEP04(2021)095}{JHEP}, 04\penalty0
  (095):\penalty0 095, 2021{\natexlab{b}}.
\newblock [arXiv:\href{http://arxiv.org/abs/2012.01889}{2012.01889}].

\bibitem[Barbero~G. et~al.(2021{\natexlab{a}})Barbero~G., D\'\i{}az,
  Margalef-Bentabol, and Villase\~nor]{concise}
J.F. Barbero~G., B.~D\'\i{}az, J.~Margalef-Bentabol, and E.J.S. Villase\~nor.
\newblock \emph{Concise symplectic formulation for tetrad gravity}.
\newblock \href{http://dx.doi.org/10.1103/PhysRevD.103.024051}{Phys. Rev. D},
  103\penalty0 (2):\penalty0 024051, 2021{\natexlab{a}}.
\newblock [arXiv:\href{http://arxiv.org/abs/2011.00661}{2011.00661}].

\bibitem[Barbero~G. et~al.(2021{\natexlab{b}})Barbero~G., D\'\i{}az,
  Margalef-Bentabol, and Villase\~nor]{diaz2021hamiltonian}
\rule[.6ex]{3em}{.05ex}.
\newblock \emph{Hamiltonian {G}otay-{N}ester-{H}inds analysis of the
  parametrized unimodular extension of the {H}olst action}.
\newblock \href{http://dx.doi.org/10.1103/PhysRevD.103.064062}{Phys. Rev. D},
  103\penalty0 (6):\penalty0 064062, 2021{\natexlab{b}}.
\newblock [arXiv:\href{http://arxiv.org/abs/2101.12311}{2101.12311}].

\bibitem[Frauendiener and Sparling(1992)]{FrauendienerSparling1992}
J.~Frauendiener and G.A.J. Sparling.
\newblock \emph{On the symplectic formalism for general relativity}.
\newblock \href{http://dx.doi.org/10.1098/rspa.1992.0010}{Proc. R. Soc. Lond.
  A}, 436:\penalty0 141–153, 1992.

\bibitem[Ashtekar and Magnon-Ashtekar(1982)]{AshtekarMagnon1982}
A.~Ashtekar and A.~Magnon-Ashtekar.
\newblock \emph{On the symplectic structure of general relativity}.
\newblock \href{http://dx.doi.org/10.1007/BF01205661}{Commun. Math. Phys.},
  86:\penalty0 55, 1982.

\bibitem[Barnich et~al.(1991)Barnich, Henneaux, and
  Schomblond]{barnich1991covariant}
G.~Barnich, M.~Henneaux, and C.~Schomblond.
\newblock \emph{Covariant description of the canonical formalism}.
\newblock \href{http://dx.doi.org/10.1103/PhysRevD.44.R939}{Phys. Rev. D},
  44\penalty0 (4):\penalty0 R939, 1991.

\bibitem[Margalef-Bentabol and Villaseñor(2021)]{CPS}
J.~Margalef-Bentabol and E.J.S. Villaseñor.
\newblock \emph{Geometric formulation of the covariant phase space methods with
  boundaries}.
\newblock
  \href{https://journals.aps.org/prd/abstract/10.1103/PhysRevD.103.025011}{\href{http://dx.doi.org/10.1103/PhysRevD.103.025011}{Phys.
  Rev. D}}, 103:\penalty0 025011, 2021.
\newblock [arXiv:\href{http://arxiv.org/abs/2008.01842}{2008.01842}].

\bibitem[Barbero~G. et~al.(2021{\natexlab{c}})Barbero~G., Margalef-Bentabol,
  Varo, and Villaseñor]{CPSGR}
J.F. Barbero~G., J.~Margalef-Bentabol, V.~Varo, and E.J.S. Villaseñor.
\newblock \emph{Covariant phase space for gravity with boundaries: {M}etric
  versus tetrad formulations}.
\newblock \href{http://dx.doi.org/10.1103/PhysRevD.104.044048}{Phys. Rev. D},
  104:\penalty0 044048, 2021{\natexlab{c}}.
\newblock [arXiv:\href{http://arxiv.org/abs/2103.06362}{2103.06362}].

\bibitem[Barbero~G. et~al.(2021{\natexlab{d}})Barbero~G., Margalef-Bentabol,
  Varo, and Villaseñor]{CPSPT}
\rule[.6ex]{3em}{.05ex}.
\newblock \emph{Palatini gravity with nonmetricity, torsion, and boundaries in
  metric and connection variables}.
\newblock \href{http://dx.doi.org/10.1103/PhysRevD.104.044046}{Phys. Rev. D},
  104:\penalty0 044046, 2021{\natexlab{d}}.
\newblock [arXiv:\href{http://arxiv.org/abs/2105.07053}{2105.07053}].

\bibitem[Barbero~G. et~al.(2022)Barbero~G., Margalef-Bentabol, Varo, and
  Villaseñor]{CPSHolst}
\rule[.6ex]{3em}{.05ex}.
\newblock \emph{On the on-shell equivalence of general relativity and {H}olst
  theories with nonmetricity, torsion, and boundaries}.
\newblock Preprint, 2022.
\newblock [arXiv:\href{http://arxiv.org/abs/2201.12141}{2201.12141}].

\bibitem[Anderson(1989)]{anderson1989variational}
I.M. Anderson.
\newblock \emph{The variational bicomplex}.
\newblock Technical report,
  \href{https://ncatlab.org/nlab/files/AndersonVariationalBicomplex.pdf}{Utah
  State Technical Report}, 1989.

\bibitem[Wald(1990)]{wald1990identically}
R.M. Wald.
\newblock \emph{On identically closed forms locally constructed from a field}.
\newblock \href{http://dx.doi.org/10.1063/1.528839}{J. Math. Phys.},
  31\penalty0 (10):\penalty0 2378, 1990.

\bibitem[Takens(1979)]{takens1979global}
F.~Takens.
\newblock \emph{A global version of the inverse problem of the calculus of
  variations}.
\newblock \href{http://dx.doi.org/10.4310/jdg/1214435235}{J. Differ. Geom.},
  14\penalty0 (4):\penalty0 543, 1979.

\bibitem[De~Le{\'o}n and Rodrigues(2011)]{de2011generalized}
M.~De~Le{\'o}n and P.R. Rodrigues.
\newblock \emph{Generalized Classical Mechanics and Field Theory}.
\newblock
  \href{https://www.elsevier.com/books/generalized-classical-mechanics-and-field-theory/de-leon/978-0-444-87753-6}{Elsevier},
  2011.

\end{thebibliography}
\end{document}